\documentclass[preprint,12pt]{elsarticle}
\usepackage{cite}

\usepackage{amssymb}

\journal{Physics A}

\begin{document}

\begin{frontmatter}


\title{Identifying significant edges via neighborhood information}
\author[1,2]{Na Zhao\fnref{kk}}
\author[1]{Jie Li\fnref{kk}}
\author[3]{Jian Wang\fnref{kk}}

\author[4]{Tong Li}
\author[1]{Yong Yu\corref{cor1}}
\ead{yuy1219@163.com}
\author[5]{Tao Zhou\corref{cor1}}
\ead{zhutou@ustc.edu}
\cortext[cor1]{Corresponding author}
\fntext[kk]{These three authors contributed equally to this work.}

\address[1]{Key Laboratory in Software Engineering of Yunnan Province, School of
Software, Yunnan University, Kunming 650091, P. R. China}
\address[2]{Electric Power Research Institute of Yunnan Power Grid, Kunming 650217,
P.R. China}
\address[3]{College of Information Engineering and Automation, Kunming University of Science and Technology, Kunming 650504, P. R. China}
\address[4]{School of Big Data, Yunnan Agricultural University, Kunming 650201, P. R. China}
\address[5]{CompleX Lab, University of Electronic Science and Technology of China, Chengdu 611471, P. R. China}





\begin{abstract}
Heterogeneous nature of real networks implies that different edges play different roles in network structure and functions, and thus to identify significant edges is of high value in both theoretical studies and practical applications. We propose the so-called second-order neighborhood (SN) index to quantify an edge's significance in a network. We compare SN index with many other benchmark methods based on 15 real networks via edge percolation. Results show that the proposed SN index outperforms other well-known methods.
\end{abstract}


\end{frontmatter}


\section{Introduction}
\label{}
Many systems in nature and human society, such as communication, social and transportation systems, can be modeled by networks \citep{albert2002statistical}. Given the heterogeneity of real networks \citep{caldarelli2007scale}, a few nodes and edges play critical roles and largely affect network structure and functions, while the majority of them are less important. Numerous methods have been proposed to identity critical nodes (see review articles \citep{Pei2013,L2016Vital} and the references therein). However, how to measure edge significance receives less attention.
\par
The identification of significant edges is highly valuable in practice. First, it can protect a system from intentional attacks. For example, one can protect a power grid from possible attacks by identifying significant transmission lines, thereby reducing the occurrences of failures. Second, deleting nodes may be too intrusive when we want to prevent failures or propagations. In comparison, edge-cutting strategies may be more applicable in some situations. For example, in financial networks, banks may withdraw certain products or reduce cooperations with some business partners to avoid financial risks, and in air transportation networks, some airlines may be closed to prevent long-range spreading of a certain disease. However, it is highly costly or even infeasible to shut down a bank or an airport. Several methods aiming at uncovering the role of a small set of significant edges in maintaining the network connectivity have been proposed \citep{holme2002attack,xia2008attack,wang2008universal,platig2013robustness,duan2016comparative}, yet we are still in need of more accurate algorithms to locate significant edges.
\par
The majority of current methods in quantifying edge significance based only on structural information. Ball \emph{et al.} \citep{ball1989finding} used the effect of removing an edge on the average distance of a network to evaluate the significance of this edge. This method is highly time-consuming and not applicable when the removal causes the network unconnected. Girvan and Newman \citep{girvan2002community} used edge betweenness (EB) to quantify the significance of an edge, which has been successfully applied in community detection. However, it requires huge computational resource. Holme \emph{et al}. \citep{holme2002attack} proposed the measure called degree product (DP), that is, the product of the degrees of two endpoints of an edge. Liu \emph{et al}. \citep{liu2015improving} proposed a measure called diffusion intensity (DI) that accounts for the function of an edge in spreading dynamics. Onnela \emph{et al}. \citep{onnela2007structure} proposed an index named topological overlap (TO), which performs very well in identifying important edges in mobile communication networks. Cheng \emph{et al}. \citep{cheng2010bridgeness} proposed the bridgeness (BN) index to detect significant edges that boost the communication between two densely connected subnetworks. Yu \emph{et al}. \citep{yu2018identifying} proposed an algorithm that combines global index of edge betweenness with local index of degree and clique. Other measures include eigenvalues \citep{restrepo2006characterizing}, link entropy \citep{qian2017quantifying}, nearest neighbor connections \citep{Ouyang2018Quantifying}, and so on.
\par
In this study, we propose the so-called second-order neighborhood (SN) index that accounts for the topological overlap of the two endpoints' second-order neighborhoods. This index can be considered as an extension of the famous TO index, and a tradeoff between computational complexity and algorithmic accuracy. To validate the significance of the edges selected by SN index, we apply the edge percolation dynamics \citep{PhysRevLett.85.5468,Moore2000Epidemics} to see whether the removal of a few edges with the highest SN values will lead to the fragmentation of the target network. Experimental results on 15 real networks demonstrate that the SN index remarkably outperforms other benchmark methods.


\section{Methods}
\label{}
Consider a connected simple network $G(V, E)$, where $V$ and $E$ are sets of nodes and edges, respectively. The number of nodes and edges are denoted by $N=|V|$ and $M=|E|$, and the edge connecting nodes $i$ and $j$ is denoted by $e_{i j}$.
 The definition of five well-known benchmarks (see more indices in the review article \citep{L2011Link}) are introduced below, followed by the description of the SN index.

EB \citep{girvan2002community} is defined as
\begin{equation}
E B(i, j)=\sum_{s \neq t} \frac{\sigma_{s t}\left(e_{i j}\right)}{\sigma_{s t}},
\end{equation}
where $\sigma_{s t}$ is the number of shortest paths from node $s$ to node $t$, and $\sigma_{s t}\left(e_{i j}\right)$  is the number of shortest paths from node $s$ to node $t$ that passing through edge $e_{i j}$. Given that EB requires the calculation of the shortest paths between all node pairs, its time complexity is $O\left(N^{3}\right)$ and thus it is unsuitable for large-scale networks.
\par
DP \citep{holme2002attack} is defined as
\begin{equation}
D P(i, j)=k_{i} k_{j},
\end{equation}
where $k_{i}$ and $k_{j}$ represent degrees of nodes $i$ and $j$, respectively.
\par
DI \citep{liu2015improving} is defined as
\begin{equation}
D I(i, j)=\frac{n_{i \backslash j}+n_{j \backslash i}}{2},
\end{equation}
where $n_{i \backslash j}$ is the number of $i$'s neighbors that are not connected to $j$ or being $j$ itself. The definition of $n_{j \backslash i}$ is similar. An illustration about how to calculate $n_{i \backslash j}$ and $n_{j \backslash i}$ is shown in figure 1.
\begin{figure}
\centering
\includegraphics[width=0.3\textwidth]{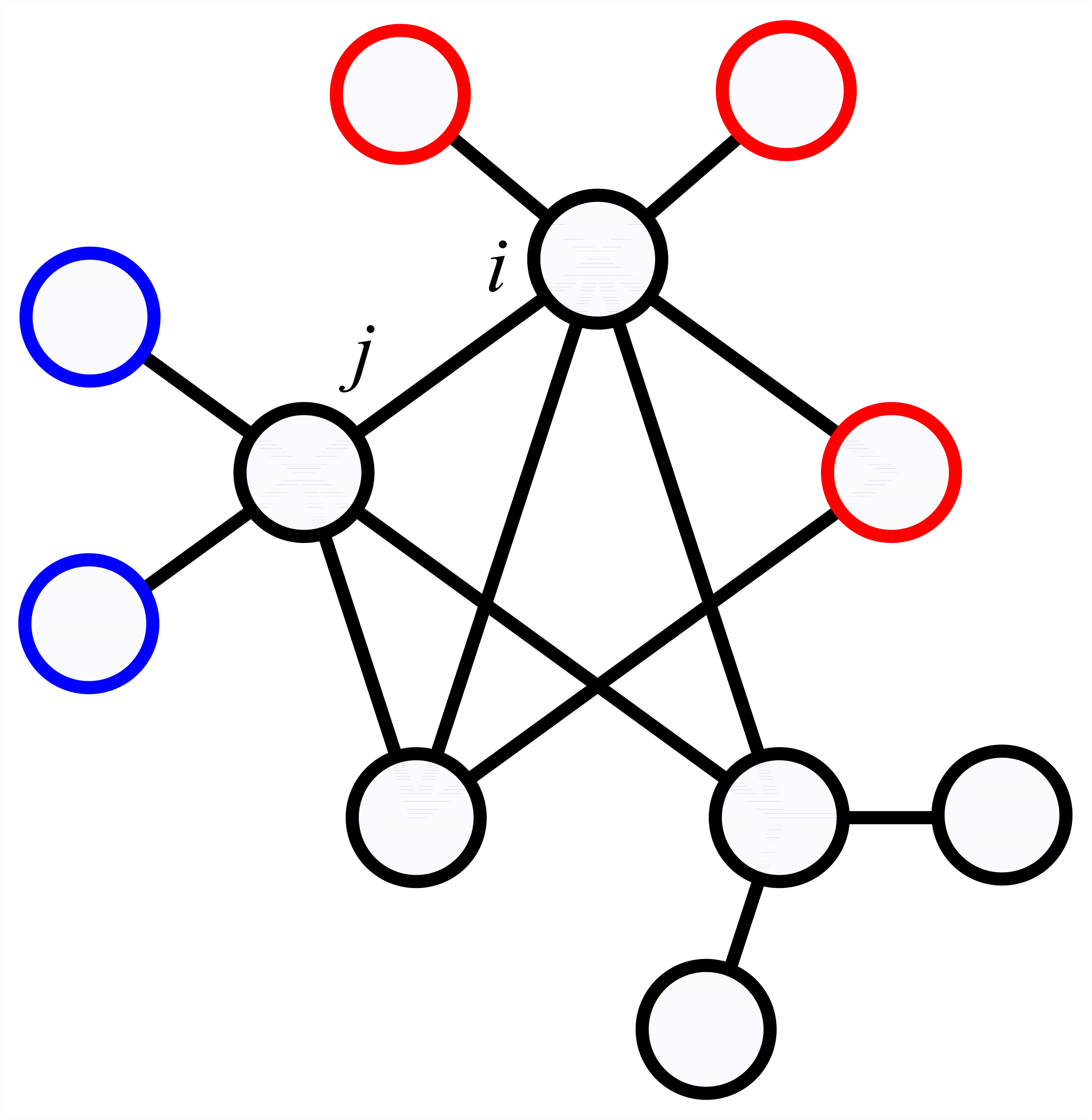}
\caption{Illustration of DI index and TO index. Since $n_{i \backslash j}$ is 3 (red nodes) and $n_{j \backslash i}$ is 2 (blue nodes), $D I(i, j)=(3+2) / 2=2.5$. As the number of common neighbors of nodes $i$ and $j$ is $n_{i j}=2$, $T O(i, j)=2 / ((6-1)+(5-1)-2) \approx 0.286$.}
\end{figure}
\par
TO \citep{onnela2007structure} is defined as
\begin{equation}
T O(i, j)=\frac{n_{i j}}{\left(k_{i}-1\right)+\left(k_{j}-1\right)-n_{i j}},
\end{equation}
where $n_{i j}$ is the number of common neighbors of nodes $i$ and $j$. This index is very similar to the Jaccard index \citep{Jaccard2010The}. A simple example about how to calculate $TO(i, j)$ is also shown in figure 1.
\par
BN \citep{cheng2010bridgeness} is defined as:
\begin{equation}
B N(i, j)=\frac{\sqrt{S_{i} S_{j}}}{S\left(e_{i j}\right)},
\end{equation}
where $S_{i}$ is the size of the largest clique containing node $i$ and $S\left(e_{i j}\right)$ is the size of the largest clique containing edge $e_{i j}$. A clique is a full connected subgraph. A simple example is illustrated in figure 2.

\begin{figure}
\centering
\includegraphics[width=0.7\textwidth]{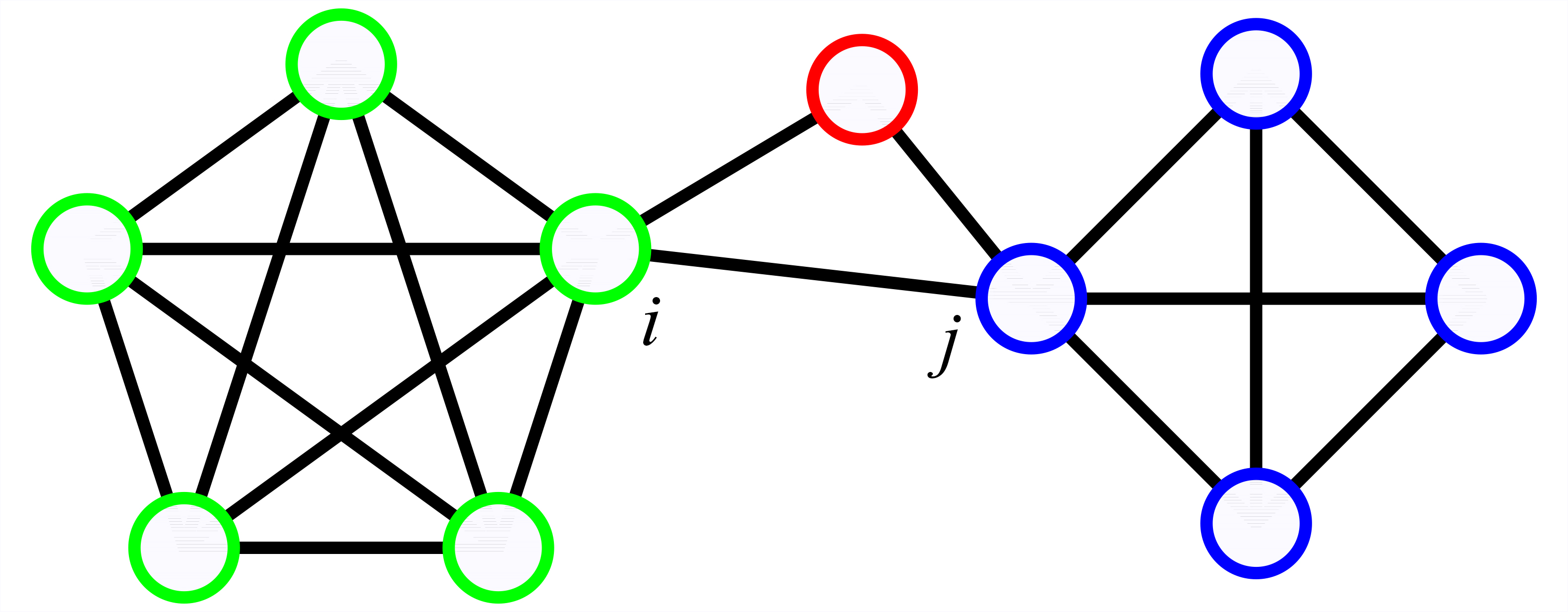}
\caption{Illustration of the BN index. As node $i$, node $j$ and edge $e_{i j}$ are contained in the corresponding largest cliques with sizes being 5, 4 and 3, $B N(i, j)=\sqrt{5 \times 4} / 3 \approx 1.49$.}
\end{figure}
\par
Next, we introduce the SN index for an arbitrary edge $e_{i j}$.
Denote $n_{i \backslash j}^{(2)}$ the number of nodes whose distance to node $i$ are 2 in the subgraph $G \backslash\left\{e_{i j}\right\}$ obtained by removing edge $e_{i j}$ from $G$. $n_{j \backslash i}^{(2)}$ is defined in a similar way.
Then, the $SN$ index is defined as:
\begin{equation}
S N(i, j)=\frac{\left|n_{i \backslash j}^{(2)} \cap n_{j \backslash i}^{(2)}\right|}{\left|n_{i \backslash j}^{(2)} \cup n_{j \backslash i}^{(2)}\right|}.
\end{equation}
\par
A simple example is illustrated in figure 3. If the second-order common neighbors of nodes $i$ and $j$ are rare, the edge $e_{i j}$ is more likely to be a potential bridge between two different communities, which is crucial in facilitating communications between these two communities \citep{Granovetter1995Getting,csermely2006weak}. Therefore, the smaller the SN index is, the more significant role the edge plays.
\begin{figure}
\centering
\includegraphics[width=0.6\textwidth]{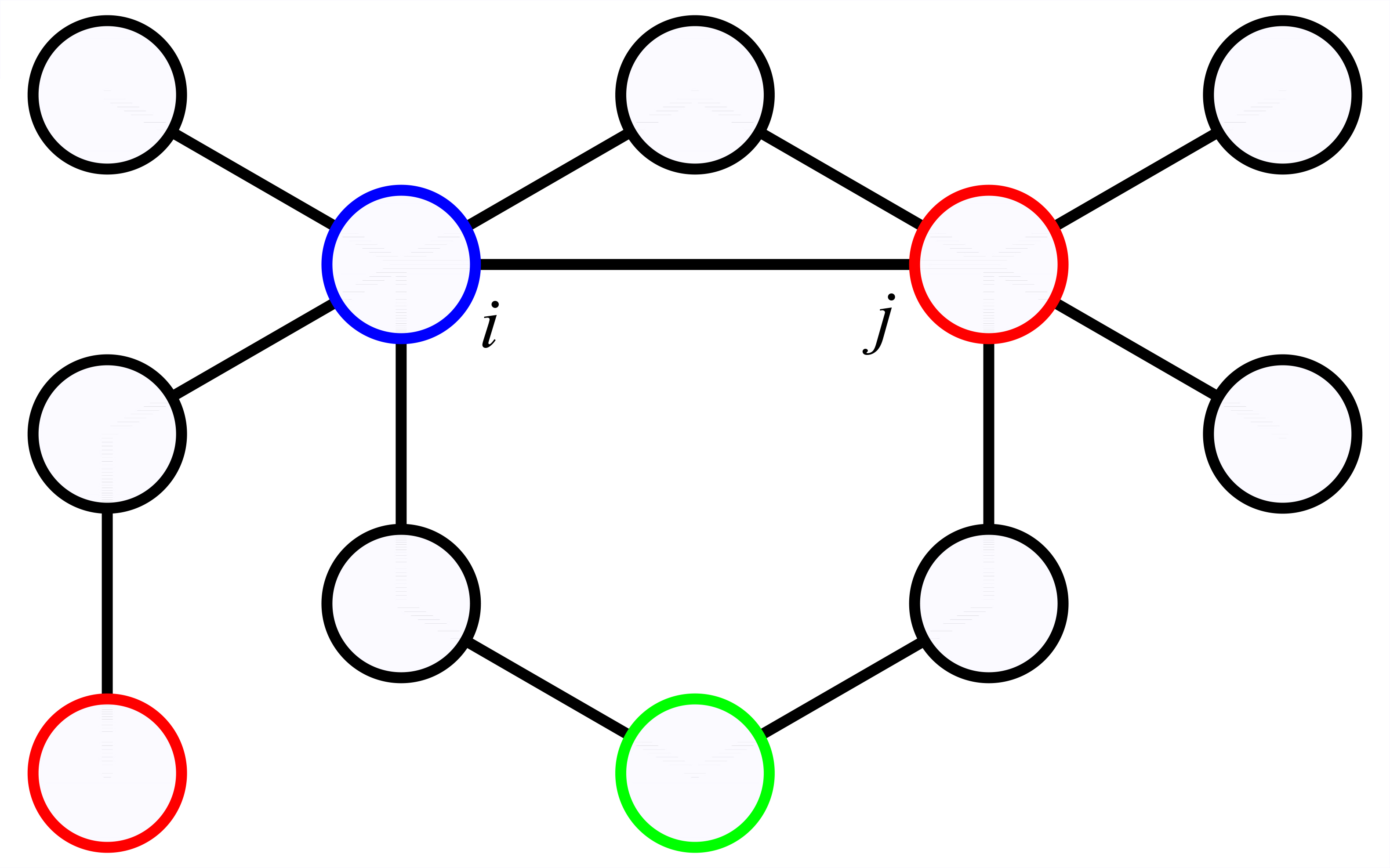}
\caption{Illustration of the SN index, where $n_{i \backslash j}^{(2)}$ is 3 (red and green nodes), $n_{j \backslash i}^{(2)}$ is 2 (blue and green nodes), and thus $S N(i, j)=1/4=0.25. $}
\end{figure}

\section{Results}
To validate the role of edges in maintaining the network connectivity, we explore the edge percolation dynamics \citep{PhysRevLett.85.5468,Moore2000Epidemics}, where in each time step an edge is removed from the target network until the remaining network becomes empty.
We employ the famous measure named robustness \citep{schneider2011mitigation} to estimate the impact of the edge removal on the network connectivity. The robustness $R$ is defined as:

\begin{equation}
R=\frac{1}{N} \sum_{\ell=1}^{N} \gamma_{\ell},
\end{equation}
where $\gamma_{\ell}$ is the ratio of the number of nodes in the maximum connected component after the removal of $\ell$ edges to the number of nodes in the original network. $1/ M$ is a normalization factor that guarantees the comparison of networks with different sizes. Obviously, a smaller value of $R$ suggests a faster fragmentation, indicating the corresponding index can better rank the edge significance.
\par
We conduct experiments on 15 real networks from different fields.
(i) \emph{Dolphins}. --- A social network with frequent associations among 62 dolphins \citep{dophinData}.
(ii) \emph{Polbooks}. --- A network of purchases of political US books \citep{polbookdata}.
(iii) \emph{Adjnoun}. --- An adjacent network of common adjectives and nouns in Charles Dickens' novel \emph{David Copperfield} \citep{adjacenciesData}.
(iv) \emph{Neural}. --- A neural network of C.elegans \citep{watts1998collective}.
(v) \emph{Jazz}. --- A collaborative network of jazz musicians \citep{jazzdata}.
(vi) \emph{Metabolic}. --- A metabolic network of C.elegans \citep{CeleganData}.
(vii) \emph{Email}. --- An e-mail exchange network among the students of the University of Rovira-Virginy \citep{emailData}.
(viii) \emph{Yeast}. --- A protein-protein interaction network of Yeast \citep{Dongbo2003Topological}.
(ix) \emph{Friendship}. --- A friendship network for users of hamsterster.com \citep{konect:2017:petster-friendships-hamster}.
(x) \emph{Kohonen}. --- A citation network from Pajek \citep{nr}.
(xi) \emph{FissionYeast}. --- A yeast fission network \citep{nr}.
(xii) \emph{Dmela}. --- A protein-protein interaction network of D.melanogaster \citep{nr,singh2008-isorank-multi}.
(xiii) \emph{Openflights}. --- The airport transportation network in Openflights.org where two airports are connected by an edge if there is at least one direct flight between them \citep{nr,opsahlanchorage}.
(xiv) \emph{Lederberg}. --- A citation network obtained from the Garfield's collection \citep{nr}.
(xv) \emph{AstroPh}. --- A collaborative network of authors having uploaded papers in the arXiv Astrophysics sector \citep{astrPhdata}.
Table 1 summarizes the basic topology characteristics of these networks.

\begin{table}
\vspace{20pt}
\centering
\begin{tabular}{cccccc}
\hline
Networks     & $N$     & $M$      & $\langle k\rangle$ & $C$      & $r$       \\ \hline
Dolphins     & 62    & 159    & 5.1290                     & 0.2590 & -0.0436 \\
Polbooks     & 105   & 441    & 8.4000                     & 0.4875 & -0.1279 \\
Adjnoun      & 112   & 425    & 7.5893                     & 0.1728 & -0.1293 \\
Neural       & 297   & 2148   & 14.4646                    & 0.2924 & -0.1632 \\
Jazz         & 198   & 2742   & 27.6970                    & 0.6175 & 0.0202  \\
Metabolic    & 453   & 2025   & 8.9404                     & 0.6465 & -0.2196 \\
Email        & 1133  & 5451   & 9.6222                     & 0.2202 & 0.0782 \\
Yeast        & 2361  & 7182   & 6.0839                     & 0.1301 & -0.0846 \\
Friendships  & 1858  & 12534  & 13.4919                    & 0.0904 & -0.0846 \\
Kohonen      & 4470  & 12731  & 5.6962                     & 0.2100 & -0.1204 \\
FissionYeast & 2031  & 25274  & 24.8882                    & 0.1874 & -0.1013 \\
Dmela        & 7393  & 25569  & 6.9171                     & 0.0118 & -0.0465 \\
Openflights  & 2918  & 30501  & 20.9054                    & 0.3967 & 0.0461  \\
Lederberg    & 8843  & 41601  & 9.4088                     & 0.2968 & -0.0996 \\
AstroPh      & 16705 & 121251 & 14.5167                    & 0.6387 & 0.2355  \\ \hline
\end{tabular}
\caption{Basic topological features of 15 real networks, where $\langle k\rangle$, $C$ and $r$ represent average degree, clustering coefficient \citep{watts1998collective} and assortativity coefficient \citep{mcpherson2001birds}, respectively.}
\label{bs}
\end{table}
\par

Figure 4 shows the collapsing processes of these real networks, resulted from the edge removal by using SN and other benchmark algorithms. Overall speaking, SN leads to much faster collapse than all other algorithms. Table 2 compares the robustness of SN and other benchmarks. Results show that, of all the benchmarks, SN index invariably produces the minimum $R$ across all the 15 real networks. That is to say, SN index perform best in identifying the most significant edges.

\begin{figure}
\centering
\includegraphics[width=0.8\textwidth]{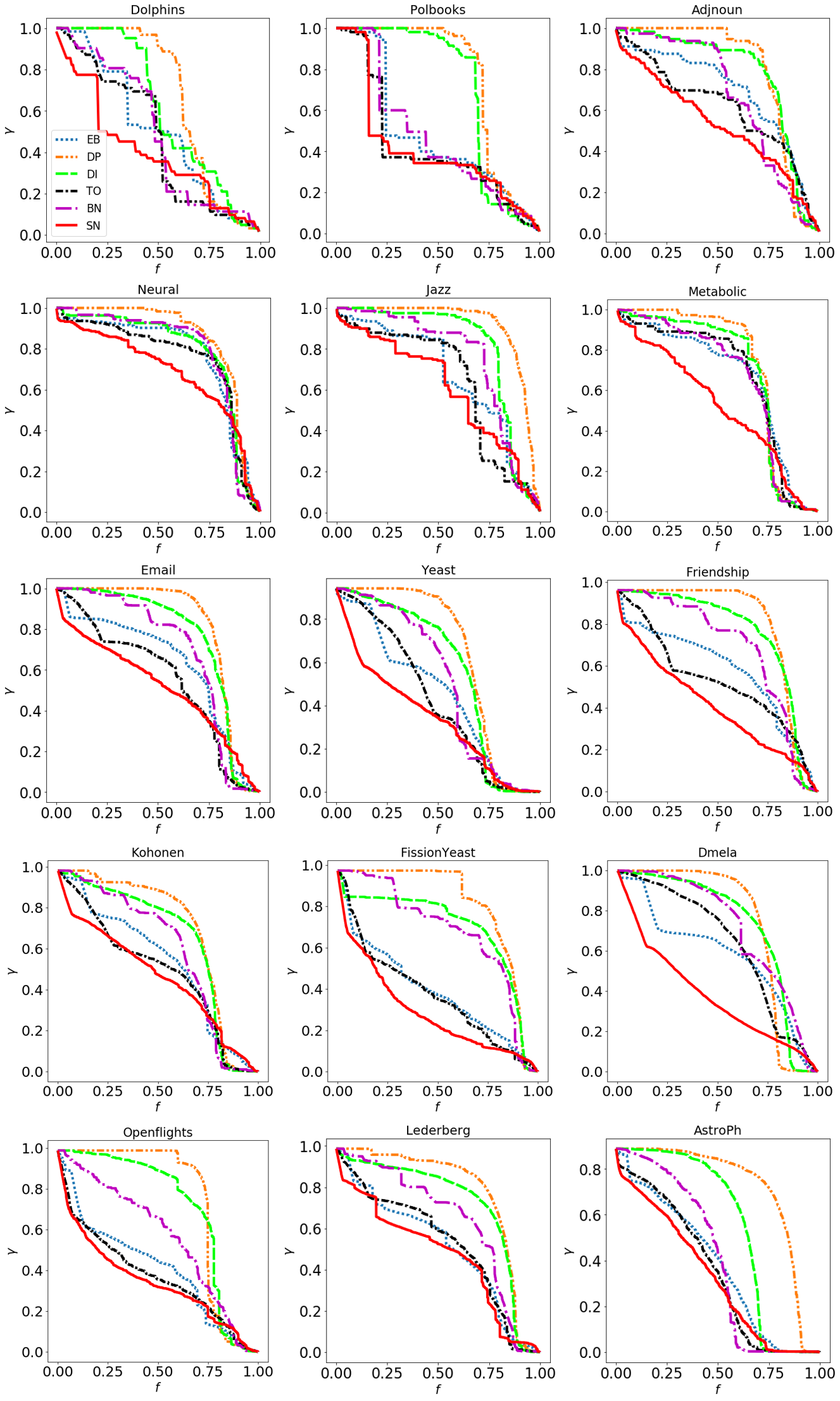}
\caption{The ratio $\gamma$ versus the fraction of edges being removed for the 15 real networks. The robustness $R$ can be simply interpreted as the area under the $\gamma-f$ curve.}
\end{figure}

\begin{table}
\centering
\begin{tabular}{ccccccc}
\hline
Networks     & EB     & DP     & DI     & TO      & BN     & SN     \\ \hline
Dolphins     & 0.5110 & 0.6746 & 0.6136 & 0.4709  & 0.4953 & \textbf{0.3991} \\
Polbooks     & 0.4773 & 0.7584 & 0.7063 & 0.4176  & 0.4921 & \textbf{0.4105} \\
Adjnoun      & 0.6841 & 0.7975 & 0.7772 & 0.6118  & 0.6642 & \textbf{0.5278} \\
Neural       & 0.6617 & 0.7270 & 0.6962 & 0.6662  & 0.6511 & \textbf{0.5216} \\
Jazz         & 0.6660 & 0.9137 & 0.8095 & 0.6398  & 0.7244 & \textbf{0.6039} \\
Metabolic    & 0.7706 & 0.8519 & 0.7903 & 0.7522  & 0.7853 & \textbf{0.6820} \\
Email        & 0.6252 & 0.8090 & 0.7621 & 0.5532  & 0.6907 & \textbf{0.5126} \\
Yeast        & 0.4493 & 0.6354 & 0.5650 & 0.4234  & 0.5100 & \textbf{0.3363} \\
Friendships  & 0.5744 & 0.7994 & 0.7598 & 0.5234  & 0.6839 & \textbf{0.4062} \\
Kohonen      & 0.5357 & 0.6946 & 0.6470 & 0.4879  & 0.5951 & \textbf{0.4616} \\
FissionYeast & 0.3882 & 0.8079 & 0.6920 & 0.3693  & 0.6641 & \textbf{0.2862} \\
Dmela        & 0.5786 & 0.7400 & 0.7300 & 0.6345  & 0.7166 & \textbf{0.3761} \\
Openflights  & 0.4114 & 0.7448 & 0.7077 & 0.3727  & 0.5661 & \textbf{0.3431} \\
Lederberg    & 0.5111 & 0.7750 & 0.7175 & 0.5244  & 0.6380 & \textbf{0.4697} \\
AstroPh      & 0.3777 & 0.7175 & 0.5521 & 0.3485  & 0.3988 & \textbf{0.3241} \\ \hline
\end{tabular}
\caption{Comparison of $R$ for different methods on 15 networks. The best result for each network (i.e., the lowest $R$ in the corresponding row) is highlighted in bold.}
\label{tab:my-table}
\end{table}

\section{Conclusion}
Identification of significant edges is of both theoretical interests and practical importance, yet it receives less attention in comparison with the challenge to dig out critical nodes. This paper provides a novel index, the SN index, to measure the importance of an edge in maintaining the network connectivity. The SN index takes into account the second-order neighborhood of each endpoint of the target edge, which can be considered as a tradeoff between indices using global topological information and indices only accounting for the nearest neighbors. It is not surprising that the SN index performs better than nearest-neighborhood-based indices like TO, while what beyond our expectation is that it performs remarkably better than the global index EB. On the basis of the extensive experiments on real networks from disparate fields, we believe that the SN index is a good candidate in quantifying an edge's significance. In the future study, one could test the performance of such index in some other dynamical processes. In addition, this work provides a simple yet clear research framework about how to identify significant edges, and thus we hope it could facilitate further studies on this issue.

\section*{Acknowledgments}
This work has been supported by the National Key Research and Development Program No. 2018YFB2100100, 2018YFC0830100, the National Natural Science Foundation of China under Grant Nos. 11975071 and 61433014. Data Driven Software Engineering innovation team of Yunnan province of China No. 2017HC012, State Key Laboratory of Computer Architecture No. CARCH201813.


\bibliographystyle{elsarticle-num-names}
\bibliography{mybib}

\begin{thebibliography}{39}
\expandafter\ifx\csname natexlab\endcsname\relax\def\natexlab#1{#1}\fi
\providecommand{\url}[1]{\texttt{#1}}
\providecommand{\href}[2]{#2}
\providecommand{\path}[1]{#1}
\providecommand{\DOIprefix}{doi:}
\providecommand{\ArXivprefix}{arXiv:}
\providecommand{\URLprefix}{URL: }
\providecommand{\Pubmedprefix}{pmid:}
\providecommand{\doi}[1]{\href{http://dx.doi.org/#1}{\path{#1}}}
\providecommand{\Pubmed}[1]{\href{pmid:#1}{\path{#1}}}
\providecommand{\bibinfo}[2]{#2}
\ifx\xfnm\relax \def\xfnm[#1]{\unskip,\space#1}\fi
\bibitem[{Albert and Barab{\'a}si(2002)}]{albert2002statistical}
\bibinfo{author}{R.~Albert}, \bibinfo{author}{A.-L. Barab{\'a}si},
\newblock \bibinfo{title}{Statistical mechanics of complex networks},
\newblock \bibinfo{journal}{Reviews of Modern Physics} \bibinfo{volume}{74}
  (\bibinfo{year}{2002}) \bibinfo{pages}{47--97}.
\bibitem[{Caldarelli(2007)}]{caldarelli2007scale}
\bibinfo{author}{G.~Caldarelli}, \bibinfo{title}{Scale-free networks: complex
  webs in nature and technology}, \bibinfo{publisher}{Oxford University Press},
  \bibinfo{year}{2007}.
\bibitem[{Pei and Makse(2013)}]{Pei2013}
\bibinfo{author}{S.~Pei}, \bibinfo{author}{H.~A. Makse},
\newblock \bibinfo{title}{Spreading dynamics in complex networks},
\newblock \bibinfo{journal}{Journal of Statistical Mechanics: Theory and
  Experiment}  (\bibinfo{year}{2013}) \bibinfo{pages}{P12002}.
\bibitem[{L{\"u} et~al.(2016)L{\"u}, Chen, Ren, Zhang, Zhang, and
  Zhou}]{L2016Vital}
\bibinfo{author}{L.~L{\"u}}, \bibinfo{author}{D.~Chen}, \bibinfo{author}{X.-L.
  Ren}, \bibinfo{author}{Q.-M. Zhang}, \bibinfo{author}{Y.-C. Zhang},
  \bibinfo{author}{T.~Zhou},
\newblock \bibinfo{title}{Vital nodes identification in complex networks},
\newblock \bibinfo{journal}{Physics Reports} \bibinfo{volume}{650}
  (\bibinfo{year}{2016}) \bibinfo{pages}{1--63}.
\bibitem[{Holme et~al.(2002)Holme, Kim, Yoon, and Han}]{holme2002attack}
\bibinfo{author}{P.~Holme}, \bibinfo{author}{B.~J. Kim}, \bibinfo{author}{C.~N.
  Yoon}, \bibinfo{author}{S.~K. Han},
\newblock \bibinfo{title}{Attack vulnerability of complex networks},
\newblock \bibinfo{journal}{Physical Review E: Statistical Nonlinear and Soft
  Matter Physics} \bibinfo{volume}{65} (\bibinfo{year}{2002})
  \bibinfo{pages}{056109}.
\bibitem[{Xia and Hill(2008)}]{xia2008attack}
\bibinfo{author}{Y.~Xia}, \bibinfo{author}{D.~J. Hill},
\newblock \bibinfo{title}{Attack vulnerability of complex communication
  networks},
\newblock \bibinfo{journal}{IEEE Transactions on Circuits and Systems II:
  Express Briefs} \bibinfo{volume}{55} (\bibinfo{year}{2008})
  \bibinfo{pages}{65--69}.
\bibitem[{Wang and Chen(2008)}]{wang2008universal}
\bibinfo{author}{W.-X. Wang}, \bibinfo{author}{G.~Chen},
\newblock \bibinfo{title}{Universal robustness characteristic of weighted
  networks against cascading failure},
\newblock \bibinfo{journal}{Physical Review E: Statistical Nonlinear and Soft
  Matter Physics} \bibinfo{volume}{77} (\bibinfo{year}{2008})
  \bibinfo{pages}{026101}.
\bibitem[{Platig et~al.(2013)Platig, Ott, and Girvan}]{platig2013robustness}
\bibinfo{author}{J.~Platig}, \bibinfo{author}{E.~Ott},
  \bibinfo{author}{M.~Girvan},
\newblock \bibinfo{title}{Robustness of network measures to link errors},
\newblock \bibinfo{journal}{Physical Review E: Statistical Nonlinear and Soft
  Matter Physics} \bibinfo{volume}{88} (\bibinfo{year}{2013})
  \bibinfo{pages}{062812}.
\bibitem[{Duan et~al.(2016)Duan, Liu, Zhou, and Ma}]{duan2016comparative}
\bibinfo{author}{B.~Duan}, \bibinfo{author}{J.~Liu}, \bibinfo{author}{M.~Zhou},
  \bibinfo{author}{L.~Ma},
\newblock \bibinfo{title}{A comparative analysis of network robustness against
  different link attacks},
\newblock \bibinfo{journal}{Physica A: Statistical Mechanics and its
  Applications} \bibinfo{volume}{448} (\bibinfo{year}{2016})
  \bibinfo{pages}{144--153}.
\bibitem[{Ball et~al.(1989)Ball, Golden, and Vohra}]{ball1989finding}
\bibinfo{author}{M.~O. Ball}, \bibinfo{author}{B.~L. Golden},
  \bibinfo{author}{R.~V. Vohra},
\newblock \bibinfo{title}{Finding the most vital arcs in a network},
\newblock \bibinfo{journal}{Operations Research Letters} \bibinfo{volume}{8}
  (\bibinfo{year}{1989}) \bibinfo{pages}{73--76}.
\bibitem[{Girvan and Newman(2002)}]{girvan2002community}
\bibinfo{author}{M.~Girvan}, \bibinfo{author}{M.~E. Newman},
\newblock \bibinfo{title}{Community structure in social and biological
  networks},
\newblock \bibinfo{journal}{Proceedings of the National Academy of Sciences of
  the United States of America} \bibinfo{volume}{99} (\bibinfo{year}{2002})
  \bibinfo{pages}{7821--7826}.
\bibitem[{Liu et~al.(2015)Liu, Tang, Zhou, and Do}]{liu2015improving}
\bibinfo{author}{Y.~Liu}, \bibinfo{author}{M.~Tang}, \bibinfo{author}{T.~Zhou},
  \bibinfo{author}{Y.~Do},
\newblock \bibinfo{title}{Improving the accuracy of the k-shell method by
  removing redundant links: From a perspective of spreading dynamics},
\newblock \bibinfo{journal}{Scientific Reports} \bibinfo{volume}{5}
  (\bibinfo{year}{2015}) \bibinfo{pages}{13172}.
\bibitem[{Onnela et~al.(2007)Onnela, Saram{\"a}ki, Hyv{\"o}nen, Szab{\'o},
  Lazer, Kaski, Kert{\'e}sz, and Barab{\'a}si}]{onnela2007structure}
\bibinfo{author}{J.-P. Onnela}, \bibinfo{author}{J.~Saram{\"a}ki},
  \bibinfo{author}{J.~Hyv{\"o}nen}, \bibinfo{author}{G.~Szab{\'o}},
  \bibinfo{author}{D.~Lazer}, \bibinfo{author}{K.~Kaski},
  \bibinfo{author}{J.~Kert{\'e}sz}, \bibinfo{author}{A.-L. Barab{\'a}si},
\newblock \bibinfo{title}{Structure and tie strengths in mobile communication
  networks},
\newblock \bibinfo{journal}{Proceedings of the National Academy of Sciences of
  the United States of America} \bibinfo{volume}{104} (\bibinfo{year}{2007})
  \bibinfo{pages}{7332--7336}.
\bibitem[{Cheng et~al.(2010)Cheng, Ren, Shen, Zhang, and
  Zhou}]{cheng2010bridgeness}
\bibinfo{author}{X.-Q. Cheng}, \bibinfo{author}{F.-X. Ren},
  \bibinfo{author}{H.-W. Shen}, \bibinfo{author}{Z.-K. Zhang},
  \bibinfo{author}{T.~Zhou},
\newblock \bibinfo{title}{Bridgeness: a local index on edge significance in
  maintaining global connectivity},
\newblock \bibinfo{journal}{Journal of Statistical Mechanics: Theory and
  Experiment}  (\bibinfo{year}{2010}) \bibinfo{pages}{P10011}.
\bibitem[{Yu et~al.(2018)Yu, Chen, and Zhao}]{yu2018identifying}
\bibinfo{author}{E.-Y. Yu}, \bibinfo{author}{D.-B. Chen},
  \bibinfo{author}{J.-Y. Zhao},
\newblock \bibinfo{title}{Identifying critical edges in complex networks},
\newblock \bibinfo{journal}{Scientific Reports} \bibinfo{volume}{8}
  (\bibinfo{year}{2018}) \bibinfo{pages}{14469}.
\bibitem[{Restrepo et~al.(2006)Restrepo, Ott, and
  Hunt}]{restrepo2006characterizing}
\bibinfo{author}{J.~G. Restrepo}, \bibinfo{author}{E.~Ott},
  \bibinfo{author}{B.~R. Hunt},
\newblock \bibinfo{title}{Characterizing the dynamical importance of network
  nodes and links},
\newblock \bibinfo{journal}{Physical Review Letters} \bibinfo{volume}{97}
  (\bibinfo{year}{2006}) \bibinfo{pages}{094102}.
\bibitem[{Qian et~al.(2017)Qian, Li, Zhang, Ma, and Lu}]{qian2017quantifying}
\bibinfo{author}{Y.~Qian}, \bibinfo{author}{Y.~Li}, \bibinfo{author}{M.~Zhang},
  \bibinfo{author}{G.~Ma}, \bibinfo{author}{F.~Lu},
\newblock \bibinfo{title}{Quantifying edge significance on maintaining global
  connectivity},
\newblock \bibinfo{journal}{Scientific Reports} \bibinfo{volume}{7}
  (\bibinfo{year}{2017}) \bibinfo{pages}{45380}.
\bibitem[{Ouyang et~al.(2018)Ouyang, Xia, Wang, Ye, Yan, and
  Tang}]{Ouyang2018Quantifying}
\bibinfo{author}{B.~Ouyang}, \bibinfo{author}{Y.~Xia},
  \bibinfo{author}{C.~Wang}, \bibinfo{author}{Q.~Ye}, \bibinfo{author}{Z.~Yan},
  \bibinfo{author}{Q.~Tang},
\newblock \bibinfo{title}{Quantifying importance of edges in networks},
\newblock \bibinfo{journal}{IEEE Transactions on Circuits and Systems II
  Express Briefs} \bibinfo{volume}{65} (\bibinfo{year}{2018})
  \bibinfo{pages}{1244--1248}.
\bibitem[{Callaway et~al.(2000)Callaway, Newman, Strogatz, and
  Watts}]{PhysRevLett.85.5468}
\bibinfo{author}{D.~S. Callaway}, \bibinfo{author}{M.~E.~J. Newman},
  \bibinfo{author}{S.~H. Strogatz}, \bibinfo{author}{D.~J. Watts},
\newblock \bibinfo{title}{Network robustness and fragility: Percolation on
  random graphs},
\newblock \bibinfo{journal}{Physical Review Letters} \bibinfo{volume}{85}
  (\bibinfo{year}{2000}) \bibinfo{pages}{5468--5471}.
\bibitem[{Moore and Newman(2000)}]{Moore2000Epidemics}
\bibinfo{author}{C.~Moore}, \bibinfo{author}{M.~E.~J. Newman},
\newblock \bibinfo{title}{Epidemics and percolation in small-world networks},
\newblock \bibinfo{journal}{Physical Review E: Statistical Physics Plasmas
  Fluids and Related Interdisciplinary Topics} \bibinfo{volume}{61}
  (\bibinfo{year}{2000}) \bibinfo{pages}{5678--5682}.
\bibitem[{L{\"u} and Zhou(2011)}]{L2011Link}
\bibinfo{author}{L.~L{\"u}}, \bibinfo{author}{T.~Zhou},
\newblock \bibinfo{title}{Link prediction in complex networks: A survey},
\newblock \bibinfo{journal}{Physica A: Statistical Mechanics and Its
  Applications} \bibinfo{volume}{390} (\bibinfo{year}{2011})
  \bibinfo{pages}{1150--1170}.
\bibitem[{Jaccard(1912)}]{Jaccard2010The}
\bibinfo{author}{P.~Jaccard},
\newblock \bibinfo{title}{The distribution of the flora in the alpine zone},
\newblock \bibinfo{journal}{New Phytologist} \bibinfo{volume}{11}
  (\bibinfo{year}{1912}) \bibinfo{pages}{37--50}.
\bibitem[{Granovetter(1995)}]{Granovetter1995Getting}
\bibinfo{author}{M.~Granovetter}, \bibinfo{title}{Getting a job: a study of
  contacts and careers}, \bibinfo{publisher}{University of Chicago Press},
  \bibinfo{year}{1995}.
\bibitem[{Csermely(2006)}]{csermely2006weak}
\bibinfo{author}{P.~Csermely}, \bibinfo{title}{Weak links: Stabilizers of
  complex systems from proteins to social networks},
  \bibinfo{publisher}{Springer Berlin}, \bibinfo{year}{2006}.
\bibitem[{Schneider et~al.(2011)Schneider, Moreira, Andrade, Havlin, and
  Herrmann}]{schneider2011mitigation}
\bibinfo{author}{C.~M. Schneider}, \bibinfo{author}{A.~A. Moreira},
  \bibinfo{author}{J.~S. Andrade}, \bibinfo{author}{S.~Havlin},
  \bibinfo{author}{H.~J. Herrmann},
\newblock \bibinfo{title}{Mitigation of malicious attacks on networks},
\newblock \bibinfo{journal}{Proceedings of the National Academy of Sciences of
  the United States of America} \bibinfo{volume}{108} (\bibinfo{year}{2011})
  \bibinfo{pages}{3838--3841}.
\bibitem[{Lusseau et~al.(2003)Lusseau, Schneider, Boisseau, Haase, Slooten, and
  Dawson}]{dophinData}
\bibinfo{author}{D.~Lusseau}, \bibinfo{author}{K.~Schneider},
  \bibinfo{author}{O.~J. Boisseau}, \bibinfo{author}{P.~Haase},
  \bibinfo{author}{E.~Slooten}, \bibinfo{author}{S.~M. Dawson},
\newblock \bibinfo{title}{The bottlenose dolphin community of doubtful sound
  features a large proportion of long-lasting associations},
\newblock \bibinfo{journal}{Behavioral Ecology and Sociobiology}
  \bibinfo{volume}{54} (\bibinfo{year}{2003}) \bibinfo{pages}{396--405}.
\bibitem[{Krebs(2004)}]{polbookdata}
\bibinfo{author}{V.~Krebs}, \bibinfo{title}{Books about {US} politics network
  dataset}, \bibinfo{year}{2004}. \URLprefix \url{http://www.orgnet.com/}.
\bibitem[{Newman(2006)}]{adjacenciesData}
\bibinfo{author}{M.~E.~J. Newman},
\newblock \bibinfo{title}{Finding community structure in networks using the
  eigenvectors of matrices},
\newblock \bibinfo{journal}{Physical Review E: Statistical Nonlinear and Soft
  Matter Physics} \bibinfo{volume}{74} (\bibinfo{year}{2006})
  \bibinfo{pages}{036104}.
\bibitem[{Watts and Strogatz(1998)}]{watts1998collective}
\bibinfo{author}{D.~J. Watts}, \bibinfo{author}{S.~H. Strogatz},
\newblock \bibinfo{title}{Collective dynamics of `small-world' networks},
\newblock \bibinfo{journal}{Nature} \bibinfo{volume}{393}
  (\bibinfo{year}{1998}) \bibinfo{pages}{440--442}.
\bibitem[{Gleiser and Danon(2003)}]{jazzdata}
\bibinfo{author}{P.~M. Gleiser}, \bibinfo{author}{L.~Danon},
\newblock \bibinfo{title}{Community structure in jazz},
\newblock \bibinfo{journal}{Advances in Complex Systems} \bibinfo{volume}{6}
  (\bibinfo{year}{2003}) \bibinfo{pages}{565--573}.
\bibitem[{Jordi and Alex(2005)}]{CeleganData}
\bibinfo{author}{D.~Jordi}, \bibinfo{author}{A.~Alex},
\newblock \bibinfo{title}{Community detection in complex networks using
  extremal optimization},
\newblock \bibinfo{journal}{Physical Review E Statistical Nonlinear and Soft
  Matter Physics} \bibinfo{volume}{72} (\bibinfo{year}{2005})
  \bibinfo{pages}{027104}.
\bibitem[{Guimera et~al.(2003)Guimera, Danon, Diaz-Guilera, Giralt, and
  Arenas}]{emailData}
\bibinfo{author}{R.~Guimera}, \bibinfo{author}{L.~Danon},
  \bibinfo{author}{A.~Diaz-Guilera}, \bibinfo{author}{F.~Giralt},
  \bibinfo{author}{A.~Arenas},
\newblock \bibinfo{title}{Self-similar community structure in a network of
  human interactions},
\newblock \bibinfo{journal}{Physical Review E: Statistical Nonlinear and Soft
  Matter Physics} \bibinfo{volume}{68} (\bibinfo{year}{2003})
  \bibinfo{pages}{065103}.
\bibitem[{Bu et~al.(2003)Bu, Zhao, Cai, Xue, Zhu, Lu, Zhang, Sun, Ling, and
  Zhang}]{Dongbo2003Topological}
\bibinfo{author}{D.~Bu}, \bibinfo{author}{Y.~Zhao}, \bibinfo{author}{L.~Cai},
  \bibinfo{author}{H.~Xue}, \bibinfo{author}{X.~Zhu}, \bibinfo{author}{H.~Lu},
  \bibinfo{author}{J.~Zhang}, \bibinfo{author}{S.~Sun},
  \bibinfo{author}{L.~Ling}, \bibinfo{author}{N.~Zhang},
\newblock \bibinfo{title}{Topological structure analysis of the protein-protein
  interaction network in budding yeast},
\newblock \bibinfo{journal}{Nucleic Acids Research} \bibinfo{volume}{31}
  (\bibinfo{year}{2003}) \bibinfo{pages}{2443--2450}.
\bibitem[{kon(2017)}]{konect:2017:petster-friendships-hamster}
\bibinfo{title}{Hamsterster friendships network dataset -- {KONECT}},
  \bibinfo{year}{2017}. \URLprefix
  \url{http://konect.uni-koblenz.de/networks/petster-friendships-hamster}.
\bibitem[{Rossi and Ahmed(2015)}]{nr}
\bibinfo{author}{R.~A. Rossi}, \bibinfo{author}{N.~K. Ahmed},
\newblock \bibinfo{title}{The network data repository with interactive graph
  analytics and visualization},
\newblock in: \bibinfo{booktitle}{Twenty-Ninth AAAI Conference on Artificial
  Intelligence}, \bibinfo{year}{2015}.
\bibitem[{Singh et~al.(2008)Singh, Xu, and Berger}]{singh2008-isorank-multi}
\bibinfo{author}{R.~Singh}, \bibinfo{author}{J.~Xu},
  \bibinfo{author}{B.~Berger},
\newblock \bibinfo{title}{Global alignment of multiple protein interaction
  networks with application to functional orthology detection},
\newblock \bibinfo{journal}{Proceedings of the National Academy of Sciences of
  the United States of America} \bibinfo{volume}{105} (\bibinfo{year}{2008})
  \bibinfo{pages}{12763--12768}.
\bibitem[{Opsahl(2011)}]{opsahlanchorage}
\bibinfo{author}{T.~Opsahl}, \bibinfo{title}{Why anchorage is not (that)
  important: Binary ties and sample selection}, \bibinfo{year}{2011}.
  \URLprefix \url{http://wp.me/poFcY-Vw}.
\bibitem[{Newman(2001)}]{astrPhdata}
\bibinfo{author}{M.~E. Newman},
\newblock \bibinfo{title}{The structure of scientific collaboration networks},
\newblock \bibinfo{journal}{Proceedings of the National Academy of Sciences of
  the United States of America} \bibinfo{volume}{98} (\bibinfo{year}{2001})
  \bibinfo{pages}{404--409}.
\bibitem[{McPherson et~al.(2001)McPherson, Smith-Lovin, and
  Cook}]{mcpherson2001birds}
\bibinfo{author}{M.~McPherson}, \bibinfo{author}{L.~Smith-Lovin},
  \bibinfo{author}{J.~M. Cook},
\newblock \bibinfo{title}{Birds of a feather: Homophily in social networks},
\newblock \bibinfo{journal}{Annual Review of Sociology} \bibinfo{volume}{27}
  (\bibinfo{year}{2001}) \bibinfo{pages}{415--444}.

\end{thebibliography}


\end{document}